\documentclass[conference]{IEEEtran}
\IEEEoverridecommandlockouts
\usepackage{cite}
\usepackage{amsmath,amssymb,amsfonts}
\usepackage{algorithm}
\usepackage{algorithmicx}
\usepackage[noend]{algpseudocode} 
\usepackage{graphicx}
\usepackage{textcomp}
\usepackage{color} 
\usepackage{xcolor}

\usepackage{caption}

\usepackage{verbatim}

\usepackage{subfigure} %
\usepackage{epstopdf} %
\usepackage{hyperref}
\usepackage{cleveref} %
\usepackage{latexsym} %
\usepackage{cancel} %

\usepackage{booktabs}

\def\BibTeX{{\rm B\kern-.05em{\sc i\kern-.025em b}\kern-.08em
    T\kern-.1667em\lower.7ex\hbox{E}\kern-.125emX}}
\begin{document}

\title{Reachability and Top-k Reachability Queries with Transfer Decay\\
}

\author{\IEEEauthorblockN{ Elena V. Strzheletska}
\IEEEauthorblockA{\textit{Department of Computer Science and Engineering,} \\
\textit{University of California, Riverside,}\\
elenas@cs.ucr.edu}
\and
\IEEEauthorblockN{Vassilis J. Tsotras}
\IEEEauthorblockA{\textit{Department of Computer Science and Engineering,} \\
\textit{University of California, Riverside,}\\
tsotras@cs.ucr.edu}

}

\maketitle

\begin{abstract}
The prevalence of location tracking systems has resulted in large volumes of spatiotemporal data generated every day. Addressing reachability queries on such datasets is important for a wide range of applications (surveillance, 
public health, social networks, etc.) A spatiotemporal reachability query identifies whether a physical item (or information etc.) could have been transferred from the source object $O_S$ to the target object $O_T$ during a time interval $I$ (either directly, or through a chain of intermediate transfers). In previous research on spatiotemporal reachability queries, the number of such transfers is not limited, and the weight of a piece of transferred information remains the same.
This paper introduces novel reachability queries, which assume a scenario of \textit{information decay}. Such queries arise when the value of information that travels through the chain of intermediate objects \textit{decreases} with each transfer. 
To address such queries efficiently over large spatiotemporal datasets, we introduce the RICCdecay algorithm. 
Further, the decay scenario leads to an important \textit{extension}: if there are many different sources of information, the aggregate value of information an object can obtain varies. As a result, we introduce a top-\textit{k} reachability problem, identifying the \textit{k} objects with the highest accumulated information. We also present the RICCtopK algorithm that can efficiently compute top-k reachability with transfer decay queries. An experimental evaluation shows the efficiency of the proposed algorithms over previous approaches.

\end{abstract}
\vspace{-0.05in}
\section{Introduction}\label{sec:introduction}
\label{introDecay}
\vspace{-0.07in}
Answering reachability queries on large spatiotemporal datasets is important for a wide range of applications, such as security monitoring, surveillance, public health, epidemiology, social networks, etc. Nowadays, with the perpetuation of Covid-19, the reachability and trajectory analysis are as important as ever, since efficient contact tracing helps to control the spread of the disease. 

Given two objects $O_S$ and $O_T$, and a time interval $I$, a spatiotemporal reachability query identifies whether information (or physical item etc.) could have been transferred from $O_S$ to $O_T$ during $I$ (typically indirectly through a chain of intermediate transfers). The time to exchange information (or physical items etc.) between objects affects the problem solution and it is application specific. An `instant exchange' scenario (where information can be instantly transferred and retransmitted between objects) is assumed in~\cite{Shahabi}, but may not be the case in many real world applications. On the other hand,~\cite{ricc} and~\cite{RICCmeet} consider two reachability scenarios without the `instant exchange' assump- tion: reachability with {\em processing delay} and {\em transfer delay}. After two objects encountered each other, the contacted object may have to spend some time to process the received information before being able to exchange it again (processing delay).
In other applications, for the transfer of information to occur, two objects are required to stay close to each other for some period of time (transfer delay); we call such elongated contact a {\em meeting}. To contract the virus, 
one has to be exposed to an infected person for a brief period of time; to exchange messages through Bluetooth, two cars have to travel closely together for some time.


While the problems discussed above considered different reachability scenarios, they had a common feature: 
the value of information carried by the object that initiated the information transmission process and the value of information obtained by any reached object was assumed to remain unchanged. In 
this paper, we remove this assumption, since for some applications it may not be valid. For example, if two persons communicate over the phone (or a Bluetooth-enabled device), some information may be lost due to faulty connection. 
We name a reachability problem, where the value of the transmitted item experiences a \textit{decay} with each transfer, as {\em reachability with transfer decay}. The formal definition of the new problem is given in Section~\ref{problemDescriptionCh4}. Note that in this paper will still assume the transfer delay scenario as this is more realistic. 




The information decay scenario leads to the second problem we introduce, namely {\em top-k reachability with decay}. Consider a group of objects (people, cars, etc...), each of which possesses a different piece of information, and starts its transmission to other objects independently of each other. The objects that initiated the process form a set of source objects. Each of the source objects may carry information of a different value (and  different weight), and during a contact, a decay of each piece of information may not be the same. As time progresses, any object may receive one or more items that originally came from different sources.
It is reasonable to compute the combined weight of all the items collected by each object and rank the objects according to their aggregate weights. Objects with the most aggregated information may be of special interest. A top-k reachability query with decay finds the $k$ objects with the highest aggregate weights.

In this paper we present two algorithms: RICCdecay and RICCtopK, that can efficiently compute reachability and top-k reachability with transfer decay queries on large spatiotemporal datasets. RICCdecay consists of two stages, preprocessing and query processing, while RICCtopK performs top-k query processing using the index from RICCdecay. 
The rest of the paper is structured as follows: Section~\ref{relWork} is an overview of related work while Section~\ref{problemDescriptionCh4} defines the two problems.
Section~\ref{preprocDecay} describes the RICCdecay algorithm and its preprocessing phase, while Sections~\ref{queryProcDecay} and \ref{queryProcTopK} present the query processing for the reachability with decay and the top-k reachability problems respectively. Section~\ref{experimentsDecay} contains the experimental evaluation and Section~\ref{conclusion} concludes the paper. 
\vspace{-0.03in}

\vspace{-0.15in}
\section{Related Work}\label{sec:introduction}
\label{relWork}

\vspace{-0.07in}
\textbf {Graph Reachability.} 
A large number of works is proposed for the static graph reachability problem.  The efficient approaches balance the preprocessing with the query processing, and are categorized in~\cite{SCARAB} as those, that use: (i) transitive closure compression~\cite{Agrawal},~\cite{dualLabl}, (ii) hop labeling~\cite{2_hop},~\cite{3_hop},~\cite{path_hop}, and (iii) refined online search~\cite{GRAIL},~\cite{preach}. 
In our model, the reachability question can be represented as a variation of a shortest path query.
The state-of-the-art algorithm for solving shortest path problems on road networks is Contraction Hierarchies (CH) ~\cite{CH}. CH benefits from creating a hierarchy of nodes on the basis of their importance for the given road network. In our problem, there is no node preference between the graph nodes, and thus applying for it CH would be inefficient. 

\indent \textbf {Evolving Graphs.} 
Evolving graphs
have recently received increased attention
The DeltaGraph~\cite{snapshot}, is an external hierarchical index structure used for efficient storing and retrieving of historical graph snapshots. For analyzing distance and reachability on temporal graphs,~\cite{tempDist} utilizes graph reachability labeling, while for large dynamic graphs,~\cite{TOL} constructs a reachability index, based on a combination of labeling, ordering, and updating techniques. These methods work with datasets of a different nature, compared with spatiotemporal. \\
\indent \textbf {Spatiotemporal Databases.}
A survey on spatiotemporal access methods appears on
in~\cite{survey_2}. They often involve some variation on hierarchical trees~\cite{Kollios:1999, Pfoser:2000,indexing,Yiu:2008,Chen:2008},
some form of a grid-based structure~\cite{Patel:2004,lugrid}, or indexing in parametric space~\cite{Ni_indexing,BHKT05}. 
The existing spatiotemporal indexes support traditional range and nearest neighbor queries and not the reachability queries we examine here. Some of the recent more complex queries were focused on querying/identifying the behavior and patterns of moving objects: discovering moving clusters~\cite{clusters_1,clusters_2}, flock patterns~\cite{flock}, and convoy queries~\cite{convoy}.

\indent {\em Spatiotemporal Reachability Queries.} 
The first disk-based solutions for the spatiotemporal
reachability problem, ReachGrid and ReachGraph appeared in~\cite{Shahabi}.  These are indexes on the contact datasets that enable faster query times. In ReachGrid, during query processing only a necessary portion of the contact network is constructed and traversed. In ReachGraph, the reachability at different scales is precomputed and then reused at query time. 
ReachGraph
makes the assumption that a contact between two objects can be instantaneous, and thus during one time instance, a chain of contacts may occur, which allows it to be smaller in size
and thus reduces query time. ReachGrid does not require the `instant exchange' assumption. 

In~\cite{ricc}, two novel types of the `no instant exchange' spatiotemporal reachability queries were introduced: reachability queries with {\em processing} and {\em transfer delays (meetings)}. The proposed solution to the first type utilized CH~\cite{CH} for {\em path contraction}. Later,~\cite{RICCmeet} considered 
the second type of delays and introduced two algorithms, RICCmeetMin and RICCmeetMax. To reduce query processing time, these algorithms precompute the shortest valid (RICCmeetMin), and the longest possible meetings (RICCmeetMax) respectively. Neither one of them can accommodate reachability queries with decay. 

\indent \textit {Spatiotemporal Top-k Queries.}
While many works have considered variations of spatial and spatiotemporal top-k  que- ries~\cite{wu,wu2011,chen,rocha,attique,ahmed,skovsgaard}, no previous work addresses the decay scenario.
\vspace{-0.2in}
\section{Problem Description}
\label{problemDescriptionCh4}
\vspace{-0.05in}
In this section, we define two novel spatiotemporal reachability problems: the problem of {\em reachability with decay} and its extension, the problem of {\em top-k reachability with decay}. 
\vspace{-0.1in}
\subsection{Background}
\vspace{-0.07in}
Let $O$ = $\{O_1, O_2, ... , O_n\}$ be a set of moving objects, whose locations are recorded for a long period of time at discrete time instants $t_1, t_2, ..., t_i, ...$, with the time interval between consecutive location recordings $\Delta t = t_{k+1} - t_k$ ($k =1, 2,...$) being constant. A {\em trajectory} of a moving object $O_i$ is a sequence of pairs $(l_i, t_k)$, where $l_i$ is the location of object $O_i$ at time $t_k$. Two objects, $O_i$ and $O_j$, that at time $t_k$ are respectively at positions $l_i$ and $l_j$, have a {\em contact} (denoted as $<O_i, O_j, t_k>$), if $dist(l_i, l_j)$ $\leq$ $d_{cont}$, where $d_{cont}$ is the {\em contact distance} (a distance threshold given by the application), and $dist(l_i, l_j)$ is the Euclidean distance between the locations of objects $O_i$ and $O_j$ at time $t_k$. 


The reachability with transfer delay scenario (which we follow here) requires to discretize the time interval between consecutive position readings $[t_k$, $t_{k+1})$ by dividing it into a series of non-overlapping subintervals $[\tau_0, \tau_1)$, ..., 
$[\tau_i$, $\tau_{i+1})$... , $[\tau_{r-1}, \tau_r)$ of equal duration $\Delta \tau = \tau_{i+1} - \tau_i$, such that $\tau_0 = t_k$ and $\tau_{r} = t_{k+1}$. We say that two objects, $O_i$ and $O_j$, had a {\em meeting}  $<O_i, O_j, I_m>$ during the time interval $I_m = [\tau_{s}, \tau_{f}]$ if they had been within the threshold distance $d_{cont}$ from each other at each time instant $\tau_{k} \in [\tau_{s}, \tau_{f}]$. The {\em duration} of this meeting is $m = \tau_{f} - \tau_{s} $. We call a meeting {\em valid} if its duration $m \geq m_q \Delta  \tau$ (where $m_q$ is the query specifies {\em required meeting duration} - time, needed for the objects to complete the exchange). Object $O_T$ is  {\em ($m_q$)-reachable} from
object $O_S$ during time interval $I = [\tau'_s, \tau'_f]$, if there exists a chain of subsequent valid meetings $<O_S, O_{i_1}, I_{m_0}>$, $<O_{i_1}, O_{i_2}, I_{m_1}>$, ... ,$<O_{i_k}, O_T, I_{m_k}>$, where each $I_{m_j} = [\tau_{s_j}, \tau_{f_j}]$ is such that $\tau_{f_j} -\tau_{s_j} \geq m_q$, $\tau'_s \leq \tau_{s_0}$, $\tau_{f_k} \leq \tau'_f$, and $\tau_{s_{j+1}} \geq \tau_{f_j}$ for $j= 0, 1, ..., k-1$. 
A reachability query determines whether object $O_T$ (the target) is reachable from object $O_S$ (the source) during time interval $I$.

\begin{figure}[h]
\vspace{-0.075in}
\centerline{\includegraphics[width=9cm, height=5.6cm]
{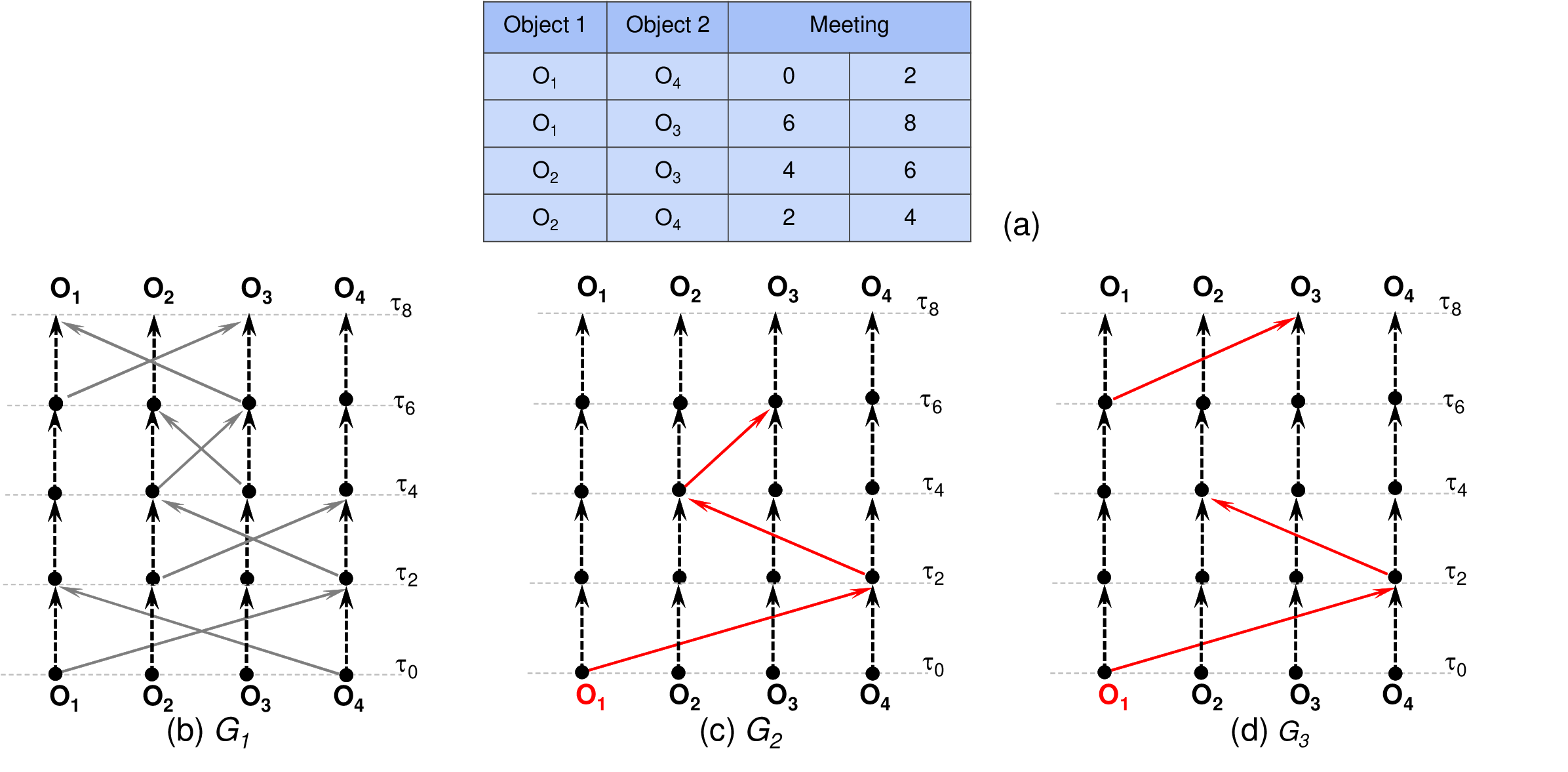}}
\vspace{-0.1in}
\caption{{(a) Record of meetings;
(b) $G_1$-meetings graph; (c) $G_2$-materialized reachability with `transfer delay' graph; (d) $G_3$-materialized reachability with `transfer decay' graph; (source object $O_1$, $m_q = 2\Delta \tau$, $d = 0.2$, $\nu = 0.6$, $I=[\tau_0, \tau_8]$).}}
  \label{MeetReachDecayTab} 
\vspace{-0.2in}
\end{figure}

Consider example in Fig.~\ref{MeetReachDecayTab}. Table $(a)$ shows the actual meetings between all objects during one time block, which is given as a meetings graph in (b). A materialized reachability graph shows how the information is being dispersed. Suppose object $O_1$ is the source object and the required meeting duration $m_q = 2\Delta \tau$. Then graph $G_2$ in (c) is the materialized ($m_q$)-reachability graph for $O_1$ on data from (a). By looking at $G_2$, one can discover all objects that can be ($m_q$)-reached by object $O_1$ during the time interval $I = [\tau_0, \tau_8]$.  


\subsection{Reachability with Decay}
\label{problemDescrDecay}
\vspace{-0.045in}
In the reachability with transfer delay scenario, to complete the transfer, it is necessary for the objects to stay within the contact distance for 
a time interval that is at least as long as the {\em required meeting duration} $m_q$. 
However, even if a meeting between objects $O_i$ and $O_j$ 
satisfied the $m_q$ requirement, under some circumstances, the transfer may still fail to occur, or the value of the transferred item may go down (e.g., a complete or partial signal loss during the communication). We consider a new type of reachability scenario, namely {\em reachability with transfer decay}, that accounts for such events. 

Let $d$ denote the {\em rate of transfer decay} - a part of information lost during one transfer ($d \in [0, 1)$). Then $p = 1 - d$ ($p \in (0, 1]$) will define the portion of the transfered information.
Suppose, the weight of the item carried by a source object $O_S$ is $w$. Then, during a valid meeting, $O_S$ can transfer this item to some object $O_i$. However, considering the decay, if $d > 0$, the value of information, obtained by $O_i$ lessens and becomes $wp$. With each further transfer, the value of the received item will continue to decrease. This process can be modeled with an exponential decay function.

We denote the number of transfers (hops), that is required to pass the information from object $O_S$ to object $O_i$ as $h$ ($h\geq0$). If $O_i$ cannot be reached by $O_S$, $h =\infty$. 
Let $g_w:\mathbb{R} \rightarrow \mathbb{R}$ be a function that calculates the weight of an item after $h$ transfers. Assuming that the transfer decay $d$ and thus $p$ are constant for the same item, $g_w(h)$ can be defined as follows: 
\vspace{-0.07in}
\begin{equation}
\label{gw}
g_w(h) = wp^h.
\vspace{-0.07in}
\end{equation}

The number of transfers $h$ in equation $($\ref{gw}$)$, that an item has to complete in order to be delivered from object $O_S$ to object $O_i$, depends on the time $\tau_j$ when it is being evaluated, and thus denoted as $h(O_i^{\tau_j})$.  Consider example in Fig.~\ref{MeetReachDecayTab}. Suppose again that $m_q = 2\Delta \tau$ and object $O_1$ is the source object. It can reach object $O_3$ by $\tau = 6$ with 3 hops, while it requires only one hop for object $O_1$ to reach $O_3$ by $\tau = 8$. So, $h(O_3^{\tau_6})$ = 3 and $h(O_3^{\tau_8})$ = 1.

The case with $p = 1$ corresponds to the reachability with transfer delay problem~\cite{RICCmeet}.
If $p < 1$, the value of $g_w(h)$ decreases exponentially with each transfer. 
Let $\nu$ denote the {\em threshold weight}. If after some transfer, the weight of the item becomes smaller than the threshold weight $\nu$, we disregard that event by assigning to the newly transferred item the weight of $0$. We say, that $h$ is the {\em allowed number of hops (transfers)} if it satisfies the threshold weight inequality  
\vspace{-0.08in}
\begin{equation}
\label{threshW}
g_w(h) \geq \nu.
\vspace{-0.07in}
\end{equation}

We denote the {\em maximum allowed number of transfers} that satisfies inequality $($\ref{threshW}$)$ as $h_{max}$. Let $f_w:\mathbb{R} \rightarrow \mathbb{R}$ be a function that assigns the weight to an item carried by object $O_i$ at time $\tau_j$, and denote it as $f_w(O_i^{(\tau_j)})$. (For brevity, we say `the weight of object $O_i$ at time $\tau_j$'.) We define $f_w(O_i^{(\tau_j)})$  as follows:

\vspace{-0.05in}
\begin{equation}
\label{fw}
f_w(O_i^{(\tau_j)}) =  \begin{cases}
g_w(h)  & \text{if $ h(O_i^{(\tau_j)}) \leq h_{max},$ }\\
0 & \text{otherwise.}\\
\end{cases}
\vspace{-0.07in}
\end{equation}



\begin{figure}[b]
\vspace{-0.2in}
\centerline{\includegraphics[width=7cm, height=6.5cm]
{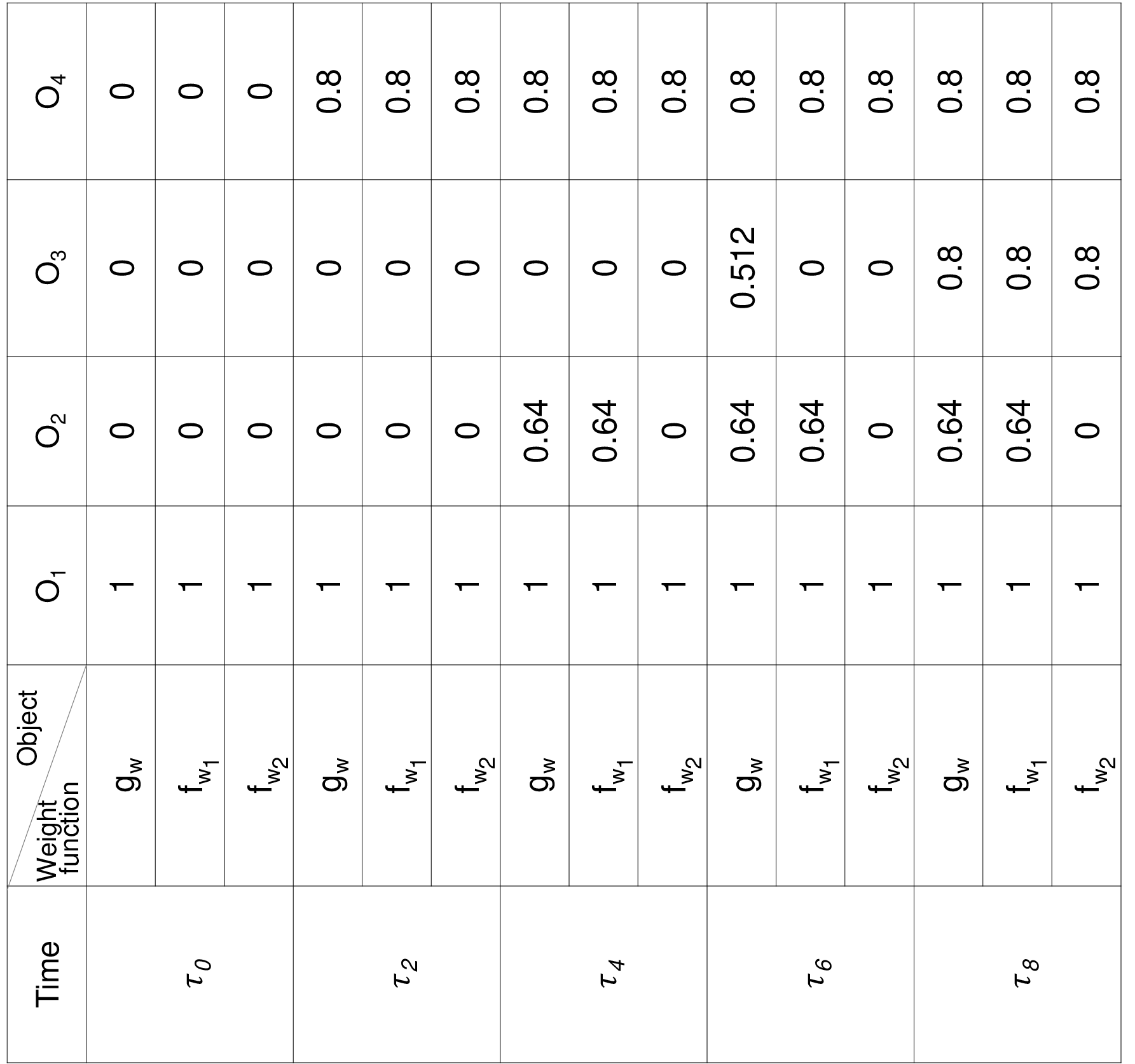}}
\caption{{The actual weight of an item $g_w$ and its assigned weights $f_{w_1}$ and $f_{w_2}$, calculated on data from Table~\ref{MeetReachDecayTab}(a) (sour- ce object $O_1, p = 0.8$, $\nu = 0.6$ for $f_{w_1}$ and $\nu = 0.7$ for $f_{w_2}$).}}

  \label{TabGwFw} 
\vspace{-0.1in}
\end{figure}

The table in Fig.~\ref{MeetReachDecayTab}(a) shows the meetings between objects $O_1, O_2, O_3,$ and $O_4$ during the time interval $I = [\tau_0; \tau_8]$. Here we assume again that $O_1$ is the source object, $m_q = 2\Delta \tau$ and $d = 0.2$ ( thus $p = 0.8$). To illustrate the difference between the actual weight of an item $g_w$ and its assigned weight $f_w$, the values $g_w$, $f_{w_1}$, and $f_{w_2}$ are computed for each object at time instants from $\tau_0$ to $\tau_8$ and recorded in the table (see Fig.~\ref{TabGwFw}). The values for the assigned weight functions $f_{w_1}$ and $f_{w_2}$ are computed for $\nu = 0.6$ and $\nu = 0.7$ respectively. The graph $G_3$ in Figure~\ref{MeetReachDecayTab}(d) is constructed for $f_{w_1}$.

Object $O_T$ is {\em ($m_q, d$)-reachable} from
object $O_S$ during time interval $I = [\tau'_s, \tau'_f]$, if there exists a chain of subsequent valid and successful (under $m_q, d$ conditions) meetings $<O_S, O_{i_1}, I_{m_0}>$, $<O_{i_1}, O_{i_2}, I_{m_1}>$, ... ,$<O_{i_k}, O_T, I_{m_k}>$, where each $I_{m_j} = [\tau_{s_j}, \tau_{f_j}]$ is such that, $\tau'_s \leq \tau_{s_0}$, $\tau_{f_k} \leq \tau'_f$, and $\tau_{s_{j+1}} \geq \tau_{f_j}$ for $j= 0, 1, ..., k-1$. The earliest time when $O_T$ can be reached is denoted as $\tau_R(O_T)$.

We assume that the values of $d$ and $\nu$ are query specified. 
An {\em ($m_q$, d)-reachability} query $Q_{md}$: $\{O_S, O_T, w, d, I, m_q, \nu\}$ determines whether the target object $O_T$ is reachable from the source object $O_S$, that caries an item whose  weight is $w$, during time interval $I = [\tau_s, \tau_f]$, given required meeting duration $m_q$, rate of transfer decay $d$, and threshold weight $\nu$, and reports the earliest time instant when $O_T$ was reached.

\vspace{-0.05in}
\subsection{Top-k Reachablility with Decay} 
\vspace{-0.05in}
\label{sec:problemDescrTopK}
We now consider the problem of top-k reachability with transfer decay.
Let $S$ = $\{O_{S_1}, O_{S_2}, ..., O_{S_q}\}$, $W = \{w_1, w_2, ..., w_q\}$, and $D = \{d_1, d_2, ..., d_q\}$ be the sets of source objects, weights, and decays respectively. Each object $O_{S_r}\in S$ carries a different piece of information (or physical item), whose weight is $w_r$, and is able to transfer this information following the ($m_q, d$)-reachability scenario. The transfer decay for the item carried by object $O_{S_r}$ is $d_r$. 

As the objects move through the network, source objects $O_{S_r}$ encounter other objects, and may pass information to them. Since each source object owns a different piece of in- formation, the transferred weight depends on both, the number of hops and the source that it came from. 
Let $h_r$ ($h_r \geq 0$) be the number of hops required for object $O_{S_r}$ to pass the information to object $O_i$. Then we can calculate the {\em actual weight} of an item $r$ after $h_r$ transfers using equation $($\ref{gw}$)$ as
\vspace{-0.04in}
\vspace{-0.075in}
$$g_{w(r)}(h_r) = w_{r}p_{r}^{h_r},$$
where $r = (1, 2, ..., q)$.
As in the previous problem, we require that each threshold weight inequality has been satisfied:
\vspace{-0.1in}
$$g_{w(r)}(h_r) \geq \nu$$ 
\vspace{-0.04in}
for $r = (1, 2, ..., q)$ and threshold weight $\nu$.

Let $h_{max(r)}$ be the maximum allowed number of transfers that satisfies the inequality above for each $r = (1, 2, ..., q)$. Similarly to $($\ref{fw}$)$, function $f_{w(r)}$ assigns weight to the $r^{th}$ item carried by object $O_i$ at time $\tau_j$ (denoted as $f_{w(r)}(O_i^{(\tau_j)})$). We define the {\em assigned weight} $f_w(O_i^{(\tau_j)})$ as follows: 
\vspace{-0.05in}
\begin{equation}
\label{fwr}
f_{w(r)}(O_i^{(\tau_j)}) =  \begin{cases}
g_{w(r)}(h_r)  & \text{if $ h_r(O_i^{(\tau_j)}) \leq h_{max(r)},$ }\\
0 & \text{otherwise.}\\
\end{cases}
\vspace{-0.02in}
\end{equation}

Furthermore, each object may receive more than one item. We denote the {\em aggregate weight} function $F_{w}:\mathbb{R} \rightarrow \mathbb{R}$ that assigns weight to the collection of items carried by object $O_i$ at time $\tau_j$ as $F_{w}(O_i^{(\tau_j)})$, and define it as follows: 
\vspace{-0.07in}
\begin{equation}
\label{Fw}
F_{w}(O_i^{(\tau_j)}) = \sum_{r = 1}^q (f_{w(r)}(O_i^{(\tau_j)})),
\vspace{-0.07in}
\end{equation}
\vspace{-0.025in}
where each $f_{w(r)}(O_i^{(\tau_j)})$ is computed as in $($\ref{fwr}$)$.
\vspace{0.07in}

A {\em top-k reachability with decay} query $Q_{topK}$ is given in the form $\{S, W, D, I, m_q, \nu, k\}$. The goal of $Q_{topK}$ is to find k objects with the highest aggregate weight $F_w$ (computed according to ~\ref{Fw}), that was obtained during the time interval $I$.

\begin{table}
\centering
  \caption{Notation used in the paper}
  \vspace{-0.05in}
  \small  
  \label{tab:notation}
  \begin{tabular}{ll}
    \toprule
    Notation & Definition\\
    \midrule
    $\Delta \tau$ & Duration between two consecutive time instants \\
    $\Delta t$ & Duration between two consecutive reporting times\\
    $m_q$, $\mu$ & Required meeting duration and  minimum meeting duration \\
    $O_S$, $O_T$ & A source and a target objects\\
    $O_i^{(\tau_j)}$ & Instance of object $O_i$ at time $\tau_j$ \\
    $\tau_R(O_i)$ & Earliest time when object $O_i$ was reached \\

    $d$, $p$ & Transfer decay and portion of transfered information \\
    $h$, $h_{max}$ & Actual and maximum allowed number of hops (transfers) \\
    $\nu$ & Threshold weight \\
    $g_w(h)$ & Actual weight of an item after $h$ transfers \\
    $f_w(O_i^{(\tau_j)})$ & Weight, assigned to an item carried by $O_i^{(\tau_j)}$ considering $\nu$ \\
    $F_w(O_i^{(\tau_j)})$ & Assigned aggregate weight of all items carried by $O_i^{(\tau_j)}$\\
    $B_k, I_k$ & Time block $k$ that spans time interval $I_k$ \\
    $C$, $H$  & Contraction parameter and grid resolution \\

    \bottomrule
\vspace{-0.1in}
  \end{tabular}
  \vspace{-0.15in}
\end{table}
\vspace{-0.07in}

\section{Preprocessing}
\label{preprocDecay}
\vspace{-0.05in}
As with other reachability problems discussed above, there are two naive approaches to solve $(m_q,d)$-reachability problem: (i) `no-preprocessing', and (ii)`precompute all'. Neither one of them is feasible for large graphs: the first does not involve any preprocessing, and thus too slow during the query processing, while the second requires too much time for preprocessing and too much space for storing the preprocessed data. To overcome the disadvantages of the second approach and still achieve fast query processing, we precompute and store only some data as described below.

In order to simplify the presentation, we assume that the minimum meeting duration $\mu$ ($\mu \leq m_q$) 
is known before the preprocessing, and set $m_q$ = $\mu$, thus fixing it. However, the proposed algorithm can be extended to work with any query specified $m_q$
by combining it with RICCmeetMax~\cite{RICCmeet}.

Suppose, our datasets contain records of objects' locations in the form  $(t, object\_ id, location)$, ordered by the location reporting time $t$. 
We start the preprocessing by dividing the time domain into a non-overlapping time intervals of equal duration ({\em time blocks}). Each time block (denoted as $B_k$) contains all records whose reporting times belong to the corresponding time period. The number of the reporting times in each block is the {\em contraction parameter} $C$. How to find an optimal value of $C$ will be discussed in Section~\ref{experimentsDecay}. 

For each time block, during the preprocessing, the following steps have to be completed: (i) computing candidate contacts, (ii) verifying contacts (performed for each $t_k$), (iii) identifying meetings, (iv) computing reachability, and (v) index construction.
Steps (i), (ii), (iii), and (v) are similar to those in~\cite{RICCmeet}; we discuss them briefly, while concentrating on step \textit{(iv)}, which is the most challenging step of preprocessing.

During the preprocessing, information regarding each object $O_i$ is saved in a data structure named {\em objectRecord}($O_i$), which is created at the beginning of each time block $B_k$ and deleted after all the needed information is written on the disk at the end of $B_k$. {\em ObjectRecord}($O_i$) has the following fields: {\em Object\_id}, {\em Cell\_id} (the object's placement in the grid with side $H$ when it was first seen during $B_k$), $ContactsRec$ (a list of the contacts of $O_i$ during $B_k$), $MeetingsRec$ (a list of meetings of $O_i$ during $B_k$). The grid side $H$ is another parameter (in addition to the contraction parameter $C$), which needs to be optimized. We will discuss this question in Section~\ref{experimentsDecay}. 
In addition, for each time block we maintain a hashing scheme, that enables to access each object's information by the object's id. 
\vspace{-0.1in}
\subsection{Computing Contacts and Finding Meetings}
\vspace{-0.05in}
Two objects $O_i$ and $O_j$ are \textit{candidate contacts} at reporting time $t_k$ if the distance between them at that time is no greater than {\em candidate contact distance} $d_{cc} = 2d_{max} +d_{cont}$ (where $d_{max}$ is the largest distance that can be covered by any object during $\Delta t$). Candidate contacts can potentially have a contact between $t_k$ and $t_{k+1}$.
To force all candidate contacts of a given object $O_i$ to be in the same or neighboring with $O_i$'s cells, at each $t_k$ we partition the area covered by the dataset into cells with side $d_{cc}$. Now, to find all candidate contacts of object $O_i$, we only need to compute the (Euclidean) distance between $O_i$ and objects in the same and neighboring cells. 

Using our assumption that between consecutive reporting times objects move linearly, at $t_{k+1}$, we can verify if there were indeed any contacts between each pair of candidate contacts during the time interval $[t_k, t_{k+1})$. If a contact occurred, it is saved in the list $ContactsRec$ of $objectRecord$ of each contacted object.
If an object $O_i$ had $O_j$ for its contact at two or more consecutive time instants, these contacts are merged into a meeting, and written in the $MeetingsRec$ list of $(O_i)$. 
At the end of each time block, a meeting duration $m$ is computed for each meeting. All meetings with
$m < \mu$ (with the exception of boundary meetings) are pruned, while all the remaining meetings are recorded into file {\em Meetings}. Boundary meetings (meetings that either start at the beginning or finish at the end of $B_k$) are recorded regardless of their duration since they may span more than one block, which needs to be verified during the query processing.

\vspace{-0.1in}
\subsection{Computing Reachability}
\vspace{-0.05in}
To speed up the query time, during the preprocessing, for each object $O_i$,
we precompute all objects that are $(\mu,d)$-reachable from $O_i$ during $B_k$. Here we are facing a challenge: to find, which objects can be $(\mu,d)$-reached by $O_i$, we need to know the transfer decay $d$ and weight threshold $\nu$, which are assumed to be unknown at the preprocessing time.  

To overcome an issue of unknown $d$ and $\nu$, we turn our problem of reachability with decay into {\em hop-reachability} problem. Recall that one of the requirements for object $O_T$ to be reachable from object $O_S$ is that each meeting in the chain of meetings from $O_S$ to $O_T$ has to be a {\em successful} meeting.

It follows from $($\ref{threshW}$)$,
that after each meeting, for each companion object $O_i$, the following condition must hold:
\vspace{-0.07in}
\setlength{\abovedisplayskip}{5pt}
$$g_w(h) = wp^h \geq \nu.$$
\vspace{-0.03in}
Thus, the allowed number of transfers (or hops) $h$ for a successful meeting should satisfy the following inequality: 
\vspace{-0.07in}
$$ h \leq \log_p \frac {\nu}{w}, $$
\vspace{-0.07in}
and finally
\vspace{-0.07in}
\begin{equation}
\label{hmax}
h_{max} = \left.\lfloor\log_p \frac {\nu}{w} \right.\rfloor.
\vspace{-0.05in}
\end{equation}

Now the problem can be stated as follows: for each object $O_i$, compute all objects, that  are ($\mu$, $h_{max}$)-reachable from $O_i$.
Moreover, for each object $O_j$ reached by $O_i$, we find the minimum number of such transfers $h_{min} \leq h_{max}$.  

Our algorithm makes use of plane sweep algorithm, where an imaginary vertical line sweeps the $xy$-plane, left-to-right, stopping at some points, where information needs to be analyzed. In our case, the x-dimension is the time-dimension, and y-dimension is the order in which the meetings are discovered.

\begin{figure}
\centerline{\includegraphics[width=9cm, height= 14cm]
{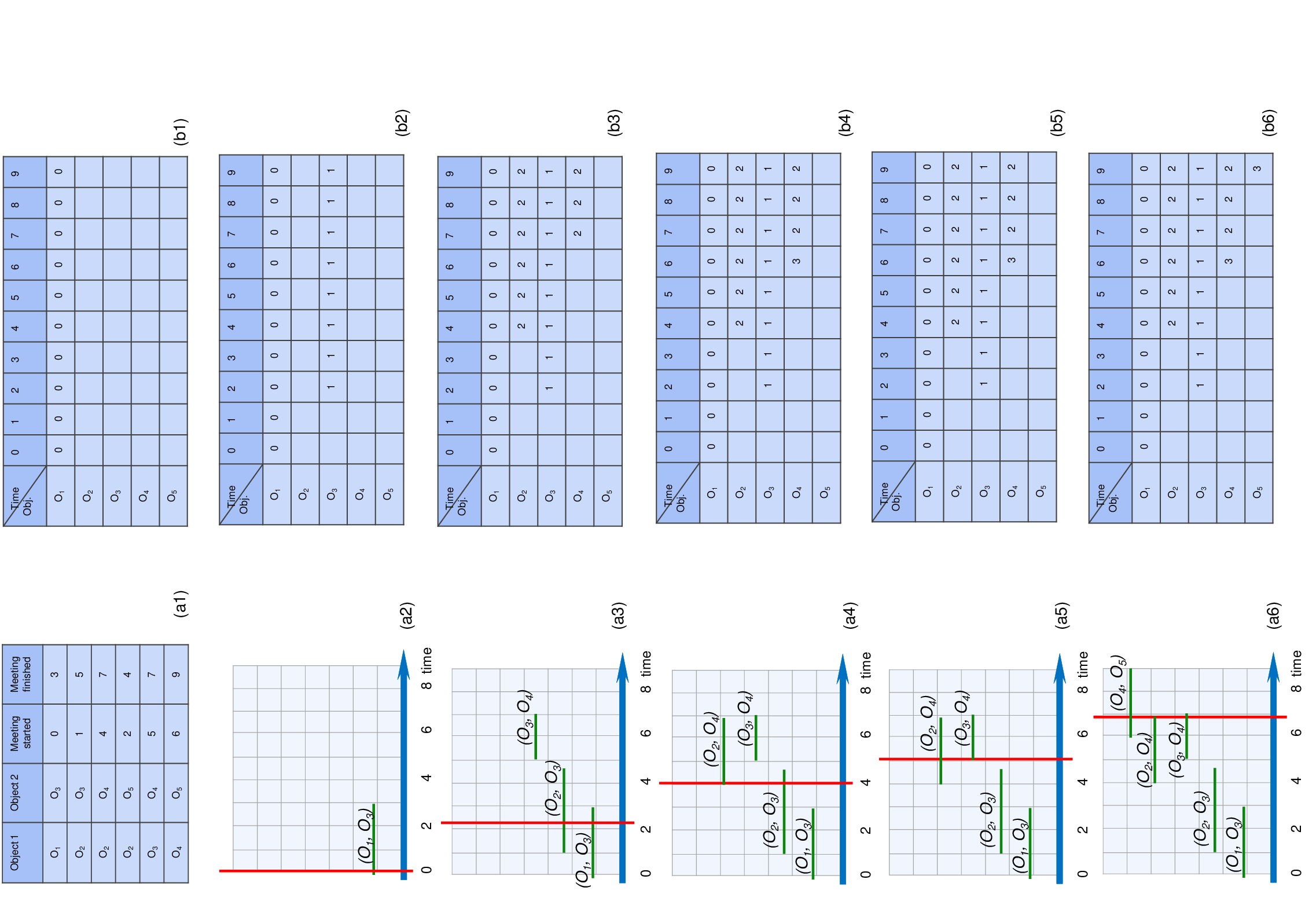}}
\vspace{-0.1in}
\caption{Computing ($h_{min}$)-reachable objects from $O_1$ ($\mu = 2$).}
  \label{decayPreproc}
\vspace{-0.2in}
\end{figure}

We demonstrate how the algorithm works on the Example in Fig.~\ref{decayPreproc}, and later provide a pseudo-code and detailed explanation. Consider the data in the table (a1). It contains records of actual meetings between all objects during one time block. (a2)-(a6) describe how reached objects and meetings are being discovered. The information about the 'reachability' status of each object is recorded into a temporary table, which is created at the beginning of each block. A row is added to the table for each reached object at the time when it is reached, and it is updated with any new event. The development of the reachability table is shown in (b1)-(b6). 

We show how to compute all objects that are reached by object $O_1$ during the given time block, assuming that $\mu = 2\Delta \tau$. At the beginning of the block, the sweep line is positioned at $\tau = 0$, and only object $O_1$ is reached (with $h_{min} = 0$), which is recorded in table (b1). During the given time block, $O_1$ has only one meeting, $<O_1, O_3, [0, 3]>$ which is placed on the plane (a2). As a result of this meeting, object $O_3$ is reached at time $\tau = 2$, with the minimum hop-value $h_{min} = 1$, which is recorded in the table (b2). 
The sweep line moves to the time $\tau = 2$ - time, when object $O_3$ was reached. Next, all meetings of $O_3$ that are either active at $\tau = 2$ or start after this time, are materialized. These are meetings $<O_3, O_2, [1, 5]>$ and $<O_3, O_4, [5, 7]>$. Consider the first meeting: $<O_3, O_2, [1, 5]>$. Even though it begins at $\tau = 1$, the retransmission does not start until $\tau = 2$, since only at this time $O_3$ becomes reached. As a result of these two meeting with object $O_3$, $O_2$ and $O_4$ become reached at $\tau = 4$ and $\tau = 7$ respectively, with $h_{min} = 2$ ((a3), (b3)). The line changes its position to $\tau = 4$. This process continues until the sweep line reaches the end of the time block. Note that the earliest reached time for an object may change, also an object's $h_{min}$ value may decrease with time. For example, object $O_4$ was reached by $O_2$ with $h_{min}=3$ at $\tau = 6$ ((a4), (b4)), however as a result of the meeting with object $O_3$, its $h_{min}$ value went down to $h_{min}=2$ at $\tau = 7$ ((a3), (b3)).


\begin{algorithm}
\vspace{-0.05in}
\caption{Reach$(h_{min})$}
\label{ReachHmin}
\medskip
\scriptsize
\label{ReachDecay}
\begin{algorithmic}[1]
\State{Input: $O_S$}
\State {procedure UpdateHmin $(O_i, \tau_s, \tau_f, h)$}
\State {\textcolor{white}{for} 
for each $\tau_k\in [\tau_s, \tau_f]$} do
{$h_{min} (O_i^{\tau_k}) = h$}
\For{each $O_i$} 
\State{$\tau_R (O_i) = \infty$}
\State{UpdateHmin$(O_i, \tau_0, \tau_{end}, \infty)$}
\Comment{$\tau_0$ and $\tau_{end}$ are the first and last time units of a block}
\EndFor
\Procedure{ReachHop}{$O_S$}
\State{$time = 0$, $\tau_R (O_S) = 0$, UpdateHmin$(O_S, \tau_0, \tau_{end}, 0)$, $S_{PQ} = \{O_S\}$},
$S_{ReachHop} = \{\emptyset\}$
\While{($(S_{PQ})$ $\neq$ $\{ \emptyset  \}$ and  time $\leq\tau_{end}$)}

\State{$O_i = $ $ExtractMin$ $(S_{PQ})$ }
\State{$S_{ReachHop} = S_{ReachHop} \cup O_i$, $time = \tau_R$ $(O_i)$}
\For{each $O_j$ that had a valid meeting with $O_i$}
\If{$O_j \notin S_{ReachHop}$}
\State{$\tau_{Rnew}(O_j) = \infty$ }
\While{$\tau_{Rnew}(O_j) \geq \tau_{R}(O_j)$}
\State{read next meeting $M_{ij} = <O_i, O_j, [\tau_s, \tau_f]>$ }
\State {compute $\tau_{Rnew}(O_j)$}
\If{$\tau_{Rnew}(O_j) < \tau_{R}(O_j)$ }
\State{\em{Update} $(S_{PQ}, O_j)$, $h=h_{min}(O_i^{time})+1$}
\If{$\tau_{R}(O_j) = \infty$} ${\tau_{R}(O_j) = \tau_{end+1}}$
\EndIf
\State{UpdateHmin($O_j, \tau_{Rnew},\tau_{R}(O_j)-1, h)$)}
\EndIf
\If{($M_{ij} = last$ $meeting <O_i, O_j>$ in $B_k$)}
\State{$\tau_{Rnew}(O_j)=-1$}
\EndIf
\EndWhile
\EndIf
\EndFor
\EndWhile
\EndProcedure 
\State{\textbf{return} $S_{Reached}$ } 
\end{algorithmic}
\end{algorithm}

\indent The process for computing all objects that are ($h_{min}$)-reachable by $O_S$ during one time block is generalized in Algorithm~\ref{ReachDecay}. Procedure UpdateHmin initializes and then updates the table that records the reachability status of each reached object.
The $S_{ReachHop}$ set keeps all objects for which all $h_{min}$ values as well as the earliest reached time had been computed and finalized. Those objects that were found to be reached, but not in $S_{ReachHop}$ yet, are placed in the priority queue $S_{PQ}$, where priority to the objects is given according to their `reached' times. When an object (say object $O_i$) that has the earliest reached time ($\tau_R(O_i)$) is extracted from $S_{PQ}$, it is placed into $S_{ReachHop}$ (lines 10, 11). At this time, all meetings of objects that can be reached by $O_i$ (but not in $S_{ReachHop}$) are analyzed (lines 13 - 23). As a result, both $\tau_R(O_j)$ (and their priority in $S_{PQ}$) as well as their $h_{min}$ values can be changed (lines 19 and 21). This algorithm has to be performed for each object of the dataset that is active during the given time block.

\begin{figure}[bh]
\vspace{-0.2in}
\centerline{\includegraphics[width=9.5cm, height= 6.5cm]
{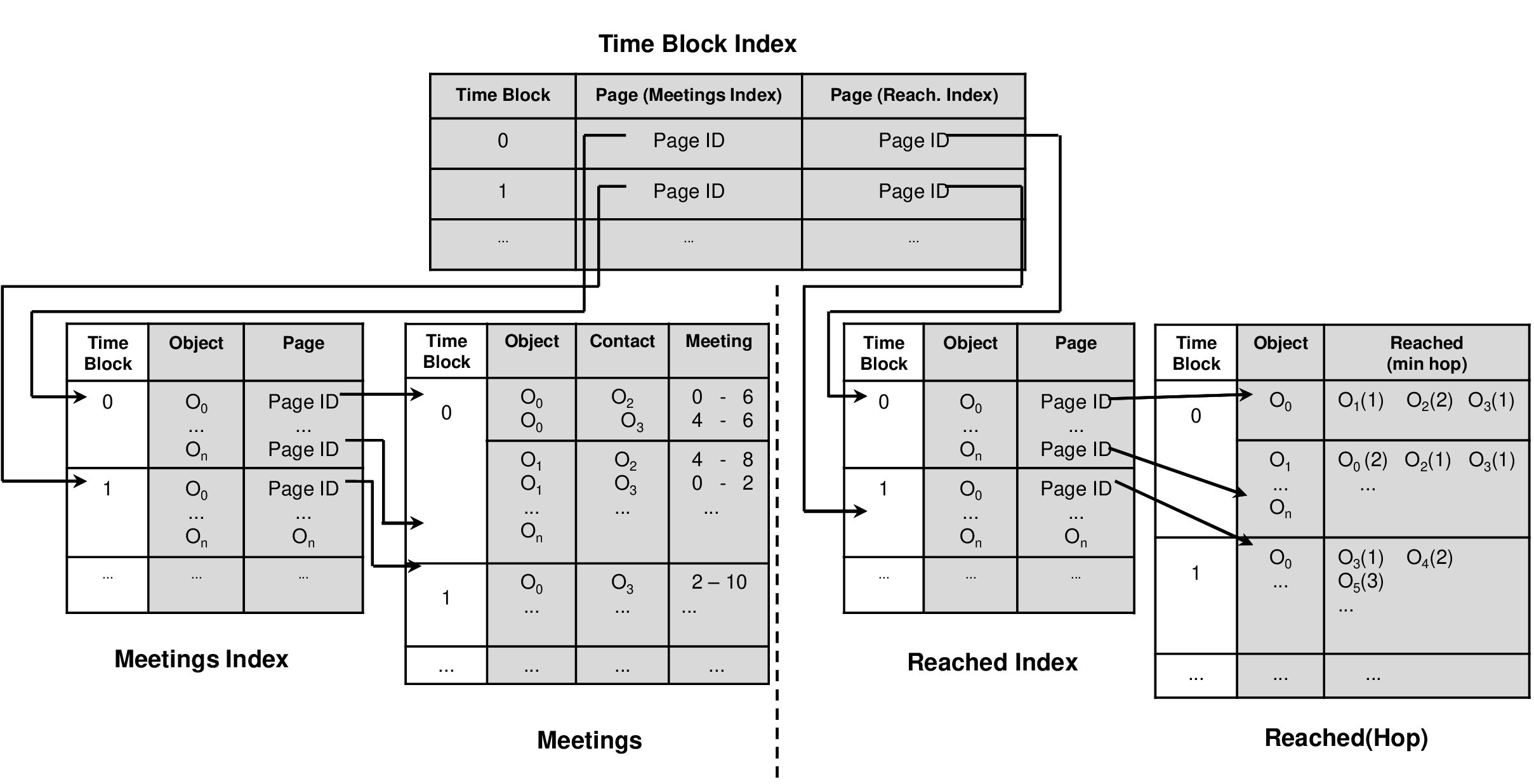}}
\vspace{-0.1in}
\caption{Two-level index on files {\em Meetings} and {\em Reached(Hop)}.}
  \label{RICCdecayIndex}
\vspace{-0.15in}
\end{figure}
\vspace{-0.1in}
\subsection{Index Construction}
\vspace{-0.05in}
The index structure of RICCdecay is similar to the one of RICCmeet algorithms~\cite{RICCmeet}: to enable an efficient search in the files {\em Meetings} and {\em Reached(Hop)} during the query processing, we create three index files: {\em Meetings Index}, {\em Reached Index}, and {\em Time Block Index} (Fig.~\ref{RICCdecayIndex}). The records in {\em Meetings Index} are organized and follow the order of time blocks. Each record contains an object id and a pointer to the page with the first record for this object (for the given time block) in file {\em Meetings}. In the {\em Reached Index}, each record consists of an object id and a pointer to the page with the first record for this object for the given time block in file {\em Reached(Hop)}. Each record in {\em Time Block Index} points to the beginning of a time block in {\em Meetings Index} and {\em Reached Index}.

\section{Reachability Queries with Decay: Query Processing}
\label{queryProcDecay}
\vspace{-0.05in}
The reachability with decay query $Q_{md}$ is issued in the form $Q_{md}$: $\{O_S, O_T, w, d,$
$[\tau_s, \tau_f], \mu,  \nu \}$. (Recall that during the preprocessing, for simplicity, we set $m_q = \mu$.) 
First, using equation $($\ref{hmax}$)$, we rewrite the problem as hop-reachability problem, replacing $w, d,$ and $\nu$ from $Q_{md}$ with $h_{max}$. The new query can be written as $Q_{mh}$: $\{O_S, O_T, h_{max}, [\tau_s, \tau_f], \mu \}$. 

The processing of $Q_{mh}$ starts from computing the time blocks $B_s$, ... , $B_f$ that contain data for the query interval $I = [\tau_s, \tau_f]$. File {\em Time Block Index} (accessed only once per query) points to the pages in the {\em Meetings Index} and {\em Reached Index} that correspond to the required blocks. These index files (accessed once per time block) in turn point to the appropriate pages in files {\em Meetings} and {\em Reached(Hop)} respectively.

The set of reached objects $S'_{reached}$ is initialized with object $O_S$ at the beginning of the query processing. We start reading file {\em Reached(Hop)} from block $B_s$, retrieving all records for object $O_S$. 
Recall that in {\em Reached(Hop)} every object $O_j$ that can be reached by object $O_i$ is recorded together with the smallest number of transfers $h_{min}$ that is required for $O_i$ to reach $O_j$. Thus during the query processing, an object $O_j$ cannot be considered as reached during the block $B_k$ unless $h_{min}(O_j^{B_k}) \leq h_{max}$ (where $h_{min}(O_j^{B_k})$ is the value $h_{min}$ of object $O_j$ at the end of $B_k$). So, each objects $O_j$ that was found to be reached by $O_S$ (a companion of $O_S$), is added to $S'_{reached}$, along with the corresponding number of hops $h_{min}$, provided that $h_{min}(O_j^{B_s}) \leq h_{max}$. Next, we proceed to block $B_{s+1}$. This time, retrieving all the companions of each object from $S'_{reached}$ and updating it by either adding new objects or adjusting the $h_{min}$ value for the objects that are already in the set. Such adjustment may be needed if, for some object $O_i \in S'_{reached}$, $h_{min}(O_j^{B_s}) > h_{min}(O_j^{B_{s+1}})$. The process continues until $O_T$ is added to $S'_{reached}$ while reading some block $B_i (i < f)$ or the last block $B_f$ is reached.

If at the end of processing $B_f$, $S'_{reached}$ does not contain the target $O_T$, the query processing can be aborted, otherwise it moves to the file {\em Meetings}. Now the process of identifying reached objects inside each block is the same as the one described in Algorithm~\ref{ReachHmin}. If there is a meeting between objects $O_i$ and $O_j$, that ends at the end of the time block, but is shorter than $m_q$, we check if it continues in the next block, and merge two meetings into one if needed. 
Also, if object $O_i$ was reached by the source object $O_S$ during the block $B_k$ with $h_{min}(O_i^{B_k})=h_1$, and in a later block $B_m$, object $O_j$ was reached by $O_i$ within $h_2$ hops, $h_{min}(O_j^{B_m})$ = $h_1+h_2$. Object $O_j$ is considered to be reached by $O_S$ if $h_{min}(O_j^{B_m}) \leq h_{max}$.  

If by the end of $B_i$, $O_T$ was not found to be reached, and $B_i < B_f$, the search switches to {\em Reached(Hop)}. This process continues until $O_T$ is confirmed to be reached by the information from {\em Meetings}, or the last block $B_f$ is processed.

\vspace{-0.05in}
\section{Top-k Reachability: Query Processing}
\label{queryProcTopK}
\vspace{-0.05in}
To process top-k reachability queries efficiently, we will use the preprocessed data and index structure from RICCdecay, described in the previous section. For that reason, we named our top-k reachability query processing algorithm {\em RICCtopK}. The top-k query $Q_{topK}$ is issued in the form  $\{S, W, D, [\tau_s, \tau_f], \mu, \nu, k\}$, where $S$ = $\{O_{S_1}, O_{S_2}, ..., O_{S_q}\}$, $W = \{w_1, w_2, ..., w_q\}$, and $D = \{d_1, d_2, ..., d_q\}$ are the sets of source objects, weights, and decays respectively. To make use of the precomputed data from RICCdecay, the top-k reachability with decay problem has to be translated into top-k hop-reachability problem. Hence, for each source object $O_{S_r} \in S$, we compute 
$h_{max(r)}$ by applying inequality (\ref{hmax}) to each triple $\{O_{S_r}, w_r, d_r\}$ as follows:
\vspace{-0.085in}
$$h_{max(r)} = \left.\lfloor \log_{p_r} \frac {\nu}{w_r} \right.\rfloor,$$
\vspace{-0.04in}
where $p_r = 1 - d_r$, and $r = \{1, 2, ... q\}.$

Now each top-k query can be thought of as written in the form $\{S, H_{hop}, [\tau_s, \tau_f], \mu, \nu,$
$k\}$, where $H_{hop}$ = $\{h_{max(1)}, h_{max(2)}, ..., h_{max(q)}\}.$ 
Note that the top-k query processing is the extension of the reachability with decay query processing algorithm, and thus we will avoid repeating some details concerning the use of the index structure during the query processing that were described earlier.


First, the set of {\em Top-k Candidates} is initialized by adding to it all source objects. We start reading file {\em Reached(Hop)} from time block $B_s$, checking all records for each source object from set $S$ (in order of their appearance in the file). Once an object, that was reached by at least one source, is discovered, it is added to {\em Top-k Candidates}. For each top-k candidate $O_i$, we keep the information about the source object(s), that it was reached by and $h_{min(r)}$ required to transfer information from each source to $O_i$. The search continues in this manner until time block $B_f$ is processed, after which the weight of each object from {\em Top-k Candidates} is computed. Note, that this is not the actual weight $F_w$ of an object, but the maximum weight $F_{max}$ that this object may receive.

Next, the query processing moves to the file {\em Meetings}. Here, the algorithm maintains two structures: {\em Top-k Candidates} and {\em Top-k}, that have to be updated at the end of each block. {\em Top-k Candidates} contains: (i) the ids of all reached objects, (ii) their corresponding maximum weights $F_{max}$, 
as well as (iii) the current weight $F_w$ of each candidate top-k object. At the beginning, the weight $F_w$ of each source object $O_{S_r}$ is set to its initial weight $w_r$, while the rest of the objects' weights $F_w$ are set to 0. 
{\em Top-k} is initialized by adding to it $k$ source objects from set $S$ with the top $k$ weights; the weight $F_w$ of each top-k object is recorded as well.

Let us denote the lowest weight $F_w$ among the objects in {\em Top-k} as $F_wmin$. If {\em Top-k} contains $k$ objects, and the object with the smallest value carries weight $F_wmin$, any object $O_i$, such that $F_{max} (O_i) < F_wmin$, cannot be among the top-k.

In file {\em Meetings}, the query processing starts from time block $B_s$. After one time block is processed, the aggregate weight $F_w$ of objects from {\em Top-k Candidates} that were involved in some transfers, may increase, and has to be updated. This may lead to changes in {\em Top-k}. After {\em Top-k} and $F_wmin$ are updated, all objects $O_i$ from {\em Top-k Candidates}, such that $F_{max} (O_i) < F_wmin$, can be removed from the set of candidates. When the work on $B_s$ is completed, we proceed to the next block. This process continues until either the last time block $B_f$ of the query is reached or the size of {\em Top-k Candidates} is reduced to the size of {\em Top-k}. The final state of {\em Top-k} answers the query.

\begin{figure*}
\vspace{-0.2in}
\centerline{\includegraphics[width=12cm, height= 10.5cm]
{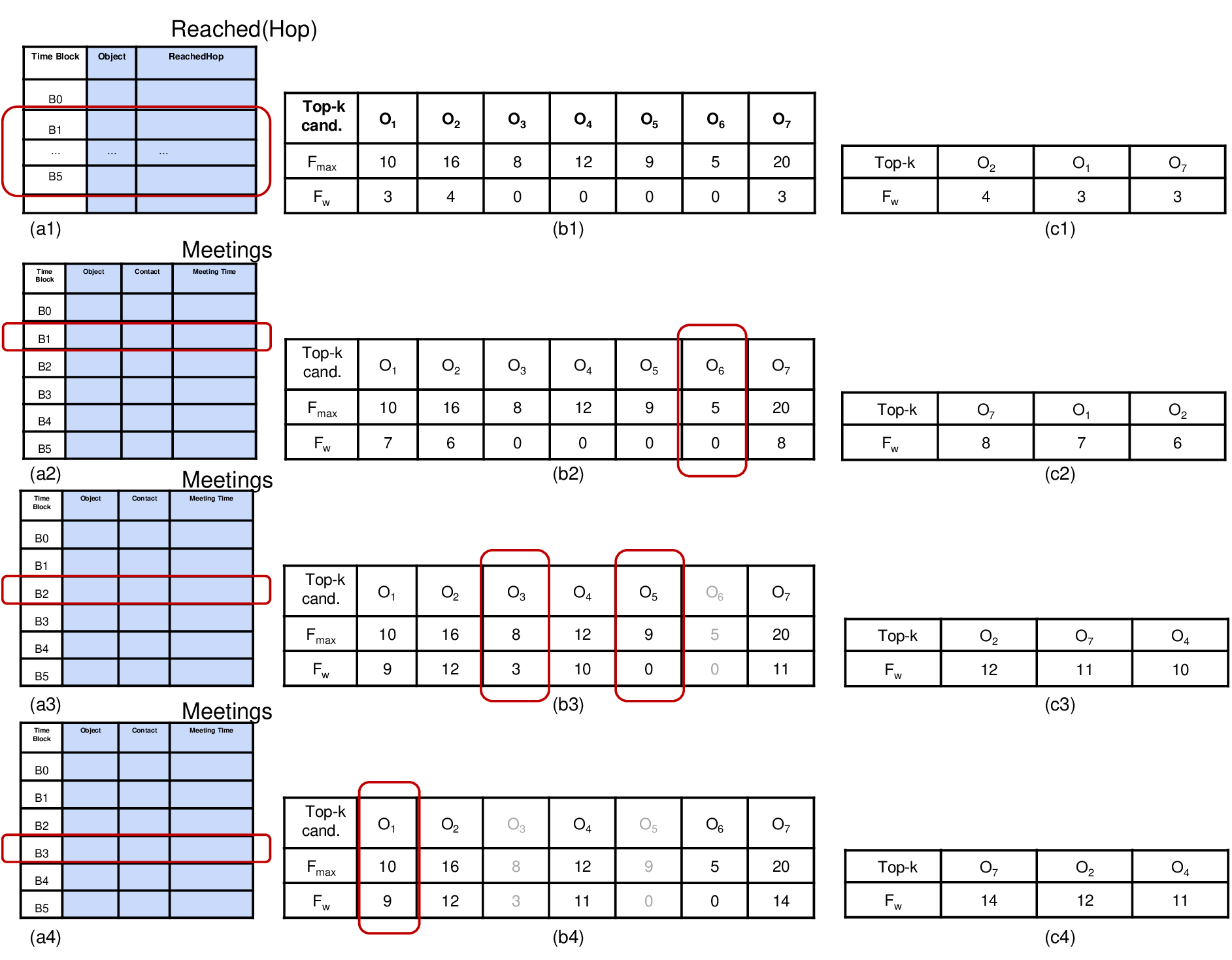}}
\vspace{-0.1in}
\caption{Top-K Query Processing (source objects: $O_1$, $O_2$, $O_7$)}
  \label{TopKAlg}
  \vspace{-0.2in}
\end{figure*}

For example, consider the top-k query with three source object $O_1$, $O_2$, and $O_7$, whose corresponding weights are 3, 4, and 3. Suppose, the query interval $[\tau_s, \tau_f]$ is contained in time blocks $B_1$ - $B_5$. Fig.~\ref{TopKAlg} illustrates the example. Fig.(a1)-(a4) show the time blocks in files {\em Reached(Hop)} and {\em Meetings} that are being processed at the given stage, tables (b1)-(b4) display the {\em Top-k Candidates} with their maximum possible aggregate weights $F_{max}$ and current aggregate weights $F_{w}$. The last column of tables, (c1)-(c4), keeps track of the current state of the {\em Top-k} set. Both, {\em Top-k Candidates} and {\em Top-k} are created after {\em Reached(Hop)} is processed and updated after the corresponding time block of file {\em Meetings} is processed.
  
The query answering begins in {\em Reached(Hop)}. The relevant data is read from blocks $B_1$ - $B_5$, and by the end of $B_5$, the superset of all objects that can be reached by the source object is identified. These objects are {\em Top-k Candidates}. They are recorded in the {\em Top-k Candidates} table, together with their maximum possible aggregate weight $F_{max}$ (b1). Since at this stage the aggregate weight $F_{w}$ is known only for the source objects, the objects $O_1$, $O_2$, and $O_7$ are placed in the {\em Top-k} (c1). The query processing moves to $B_1$ in file {\em Meetings} (a2). At the end of $B_1$, the aggregate weight of some objects $F_{w}$ is updated, and thus both, {\em Top-k Candidates} and {\em Top-k} are updated as well ((b2), (c2)). We notice that
$F_wmin(O_2)$ = 6. Thus all objects $O_i$ with
$F_{max}(O_i) < 6$ can be removed from the set of candidates. (Such objects are shown in gray in (b3) and (b4).) The next block is $B_2$ (a3), and after updating both tables ((b3) and (c3)), we exclude $O_3$ and $O_5$ from further consideration. After processing $B_3$, we remove $O_1$ from {\em Top-k Candidates}. Even though, the query interval ends only in $B_5$, the query can be suspended as the size of {\em Top-k Candidates} is reduced to the size of {\em Top-k}. 

  \vspace{-0.05in}
\section{Experimental Evaluation}
\label{experimentsDecay}
  \vspace{-0.05in}
We proceed with the results of the experimental evaluation of RICCdecay and RICCtopK. Since there are no other algorithms for processing spatiotemporal reachability queries with decay, we compare against a modified version of RICCmeetMin ~\cite{RICCmeet} that enables it to answer such queries. 
All experiments were performed on a system running Linux with a 3.4GHz Intel CPU, 16 GB RAM, 3TB disk and 4K page size. All programs were written on C++ and compiled using gcc version $4.8.5$ with optimization level $3$.
\vspace{-0.1in}
\subsection{Datasets}
\vspace{-0.05in}
All experiments were performed on six realistic datasets of two types: 
Moving Vehicles (MV) and Random Walk (RW). The MV datasets were created by the Brinkhoff data generator~\cite{brinkhoff}, which generates traces of objects, moving on real road networks. For the underlying network we used the San Francisco Bay road network, which covers an area of about $30000$ $km^2$. These sets contain information about $1000, 2000$, and $4000$ vehicles respectively (denoted as $MV_1$, $MV_2$, and $MV_4$). The location of each vehicle is recorded 
every $\Delta t = 5$ seconds during 4 months, which results in $2,040,000$ records for each object. The size of each dataset (in GB) appears in Table~\ref{AuxDataDecay}. For the experiments on these sets, $d_{cont} = 100$ meters (for a (class 1) Bluetooth connection).

\vspace{-0.3em}
\begin{table}[h]
\centering
 \caption{Size of datasets, auxiliary files and indexes}
\begin{minipage}[b]{0.3\textwidth}
\centerline{\includegraphics[width=8cm, height=2cm]{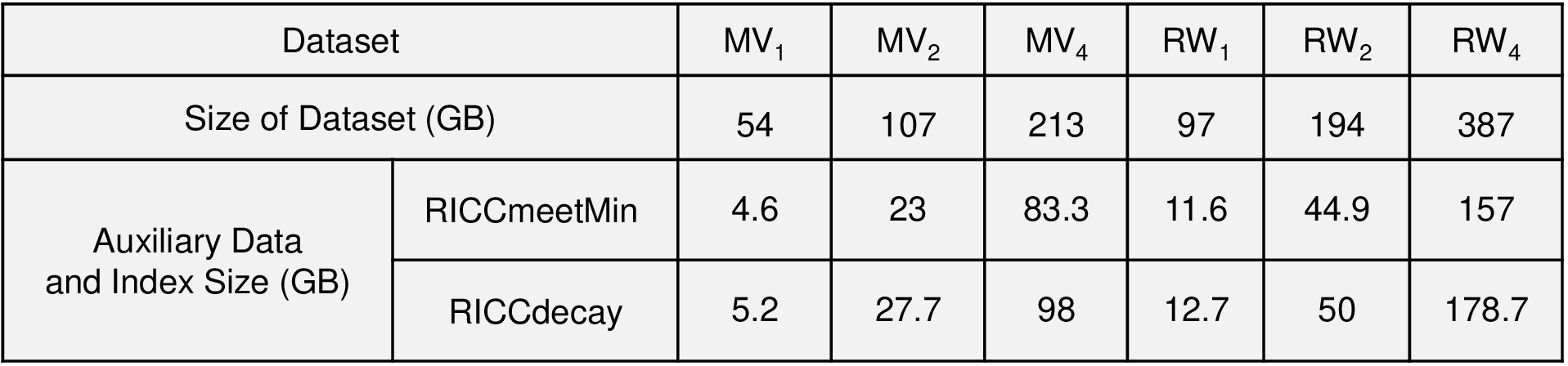}}
 \label{AuxDataDecay}
\end{minipage}
\end{table}



For the RW datasets, we created our own generator, which utilizes the modified random waypoint model~\cite{RWM}, and is often used for modeling movements of mobile users. In our model, $90\%$ of individuals are moving, while the remaining $10\%$ are stationary. At the beginning of the first trip, each user chooses whether to move or not (in the ratio of $9:1$). Each out of $90\%$ moving users chooses the direction, speed (between $1.5m/s$ and $4m/s$), and duration of the next trip, and then completes it. At the end of the trip, each person determines the parameters for the next trip, and so on. RW datasets consist of trajectories of $10000, 20000,$ and $40000$ individuals respectively (denoted as {$RW_1,RW_2,$} and {$RW_4$}). Each set covers an area of $100$ $km^2$. The location of each user is recorded every $\Delta t = 6$ sec for a period of one month (432,000 records for each person). We set $d_{cont} = 10$ meters (to identify physical contacts or contacts in the range of a Bluetooth-enabled devices).

The performance was evaluated in terms of disk I/Os during query processing. The ratio of a sequential I/O to a random I/O is system dependent; for our experiments this ratio is 20:1 (20 sequential I/Os take the same time as 1 random). 
We thus present the equivalent number of random I/Os using this ratio.

\vspace{-0.05in}
\subsection{Parameter Optimization}
\vspace{-0.05in}
The values of the contraction parameter $C$ and the grid resolution $H$, that are used for the preprocessing, depend on the datasets. For each dataset, the parameters $C$ and $H$ were tuned on the $5\%$ subset as follows. We performed the pre- processing of each subset for different values of $(C, H)$, and tested the performance of RICCdecay on a set of $200$ queries. The length of each query was picked uniformly at random bet- ween 500 and 3500 sec for the MV, and between 600 and 4200 sec for the RW datasets. The $h_{max}$ value was picked uniformly at random from 1 to 4 (we stopped at $h_{max} = 4$ since the higher the $h_{max}$, the less information is caried by the reached object and thus presents less interest). The parameters $C$ and $H$ were varied as follows: grid resolution $H$ - from 500 to 40000 meters for MV datasets, and from 250 to 2000 m for RW datasets; contraction parameter $C$ - from $0.5$ to $30$ min. For each dataset, the pair $(C, H)$ that minimized the number of I/Os was used for the rest of the experiments. For example, for $MV_1$ we used $C = 14$ min and $H = 20000$ m, while for $RW_4$ we used $C = 2$ min nd $H = 500$ m. 
\vspace{-0.05in}
\subsection{Preprocessing Space and Time}
\vspace{-0.05in}
The sizes of the auxiliary files and the index sizes for RICCmeetMin and RICCdecay appear in Table~\ref{AuxDataDecay}. RICCdecay uses about 13.5\%  more space compare to RICCmeetMin since it records more information into the file {\em Reached(Hop)}. (For each reached object, in addition to its id, it saves its hop value.) 
The time needed to preprocess one hour of data for RICCdecay ranges from 14 sec for $MV_1$ to $91$ min for $RW_4$. For comparison, the preprocessing time for RICCmeetMin ranges from 13 sec for $MV_1$ to $56$ min for $RW_4$.

 
\vspace{-0.05in}
\subsection{Query Processing}
\vspace{-0.05in}
The performance of RICCdecay was tested on sets of $100$  queries of different time intervals 
and various $h_{max} = 1, 2, 3, 4$, while $\mu$ was set to 2 sec, and the initial weight $w$ of the item carried by $O_S$ was set to $1$ for all the experiments.

\begin{figure}[h]
 \vspace{-0.15in}
\centerline{\includegraphics[width=9.3cm, height=5.25cm]
{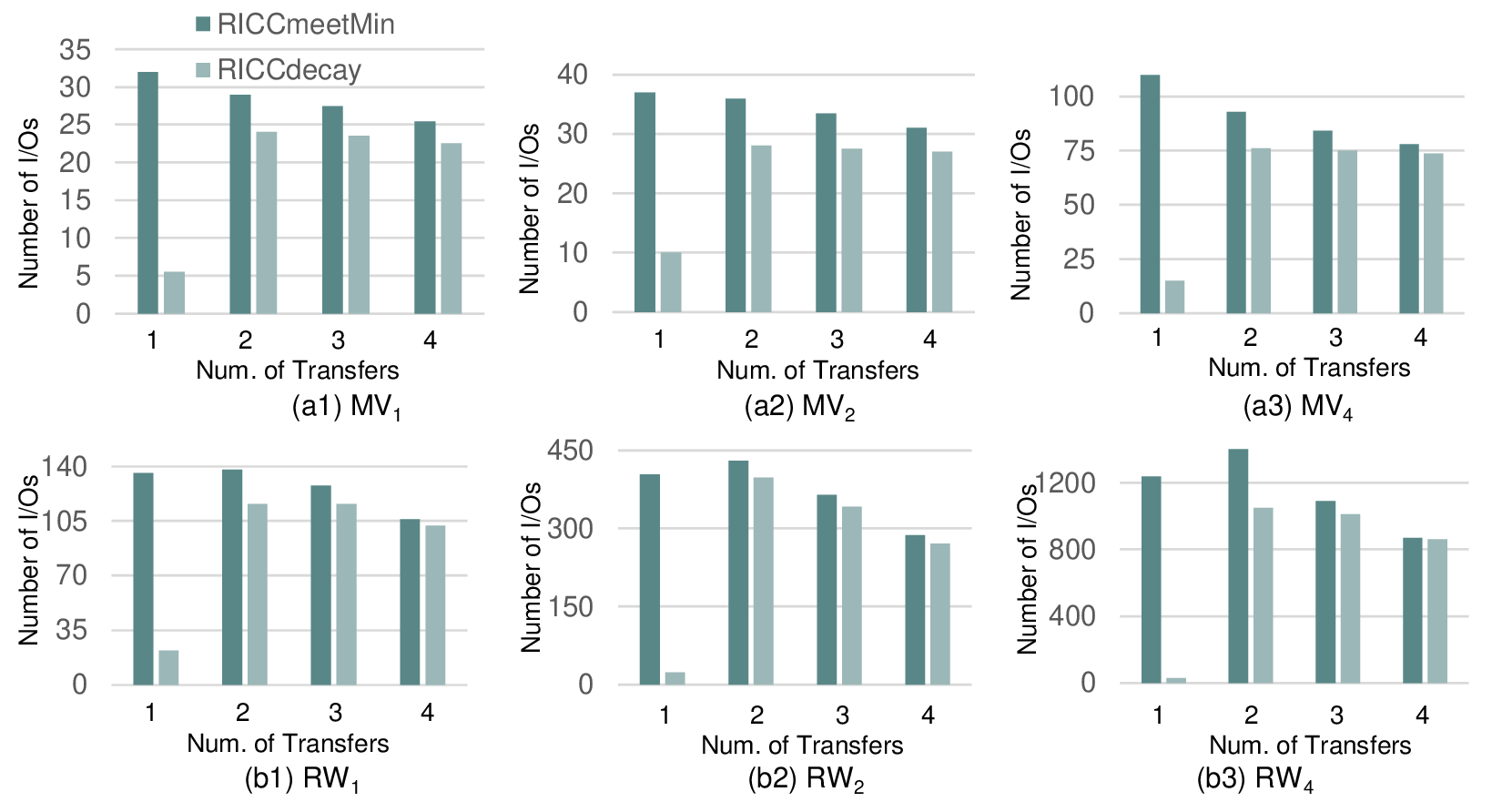}}
 \vspace{-0.1in}
\caption{Increasing maximum allowed number of transfers}
  \label{IncreaseHop}
 \vspace{-0.1in}
\end{figure}

{\bf\em Increasing the Maximum Allowed Number of Transfers.}
In this set of experiments, we analyze the impact of $h_{max}$ on the performance of the RICCdecay, and compare RICCdecay with RICCmeetMin. 
(The last was modified to enable it to answer reachability queries with decay.) We ran a set of 100 queries varying $h_{max}$ from $1$ to $4$; each query's interval was picked uniformly at random from $500$ to $3500$ sec for the MV datasets, and from $600$ to $4200$ sec for RW datasets. The results are presented in Fig.~\ref{IncreaseHop} ($a1-b3$). 
RICCdecay accesses from 1.8 (for $MV2$ dataset) to 11.5 (for $RW4$ dataset) times less pages than RICCmeetMin. The biggest advantage of RICCdecay is achieved for $h_{max} = 1$ for all datasets, and in general, the smaller the $h_{max}$, the better is the performance of the RICCdecay algorithm. When answering a query $Q_{mh}$, it reads file {\em Reached(Hop)} first. File {\em Meetings} needs to be read only if during traversing file {\em Reached(Hop)}, the target object appears among the objects, reached by the source (i.e. if $O_T\in S'_{Reached}$). However, $S'_{Reached}$ is a superset of the set of objects that can be reached by $O_S$ during the query interval $I$. We say that a query is {\em pruned}, if it aborts after reading file {\em Reached(Hop)} because of not finding the target among the reached objects. By precomputing the hop value of each reached object, {\em Reached(Hop)} gives more accurate information, than RICCmeet, which reduces the size of $S'_{Reached}$. The smaller the $h_{max}$, the less objects are in $S'_{Reached}$, and thus the higher percent of queries can be pruned. 

\begin{figure}
\centerline{\includegraphics[width=9.3cm, height= 5.25cm]
{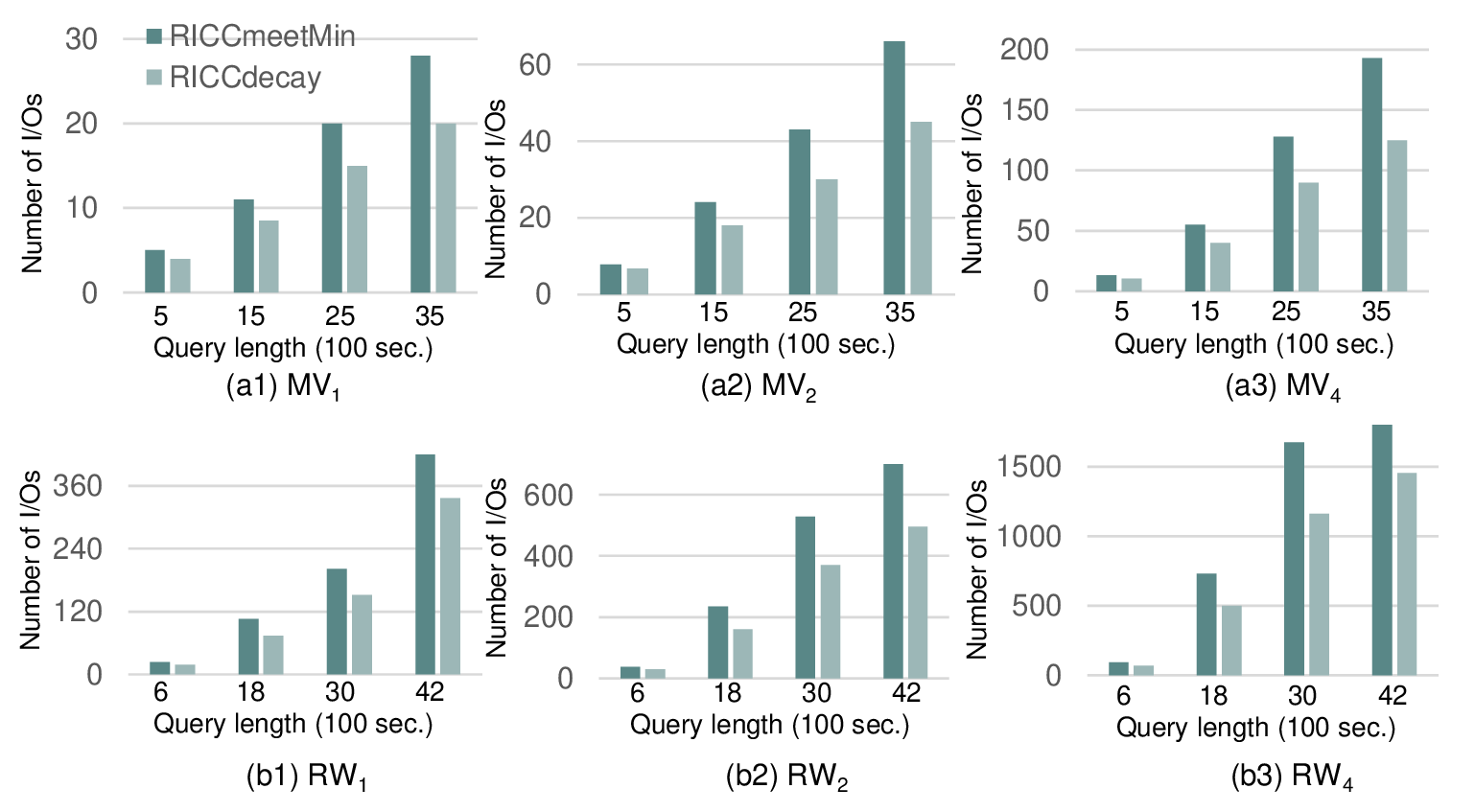}}
\vspace{-0.1in}
\caption{Increasing query length}
  \label{IncreaseQLen} 
 \vspace{-0.1in}
\vspace{-0.1in}
\end{figure}
{\bf\em Increasing Query Length.}
Now we test the performance of RICCdecay for various query lengths and compare with that of RICCmeetMin. Each test was run on a set of $100$ queries varying query length from $500$ to $3500$ sec for $MV$, and from $600$ to $4200$ sec for $RW$ datasets. The $h_{max}$ value for each query was picked uniformly at random from $1$ to $4$. The results are shown in Fig.~\ref{IncreaseQLen}. For these sets of queries, RICCdecay outperforms RICCmeetMin in all the tests, accessing about 44\% less pages in average, and this result does not change significantly from one dataset to another. 
 
{\bf\em Top-K Reachability Queries.} All the queries considered in this section until now were one-to-one queries: they had one source and one target object. Top-k queries that we described in Section~\ref{problemDescriptionCh4}
may have more than one source and one target objects. Multiple sources lead to the increase in the search space, while multiple undefined targets prohibit from the early query suspension.
In addition, the need to calculate and compare the aggregate weights of the reached objects makes it impossible to prune a query (suspend it after just searching the file {\em Reached(Hop)}). 

\begin{figure}[htb]
\vspace{-0.2in}
\centerline{\includegraphics[width=7.5cm, height= 2.5cm]
{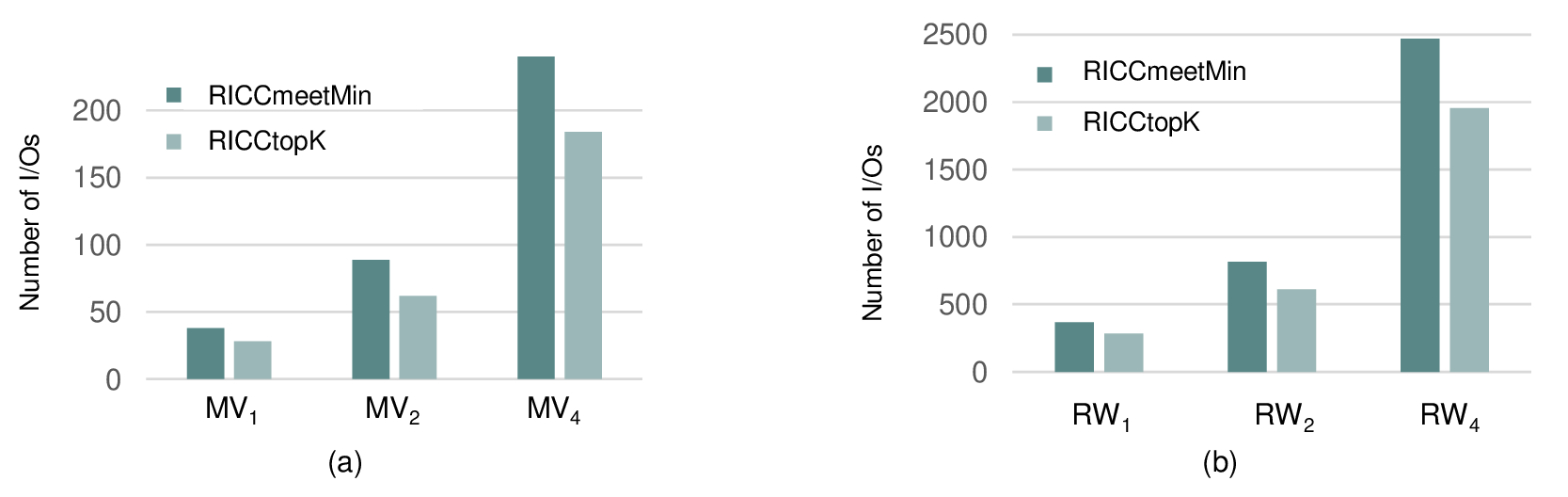}}
\vspace{-0.15in}
\caption{Top-k reachablility queries}
  \label{TopKExper} 
 \vspace{-0.25in}
\end{figure}

For each of our top-k experiments, we used sets of 100 queries, where query length was $3500$ sec for $MV$ datasets and $4200$ sec for $RW$ datasets. The number of source objects was set $4$: $S$ = $\{O_{S_1}, O_{S_2}, O_{S_3}, O_{S_4}\}$, and each weight was assigned a value of 1. Further, $D = \{0.10, 0.15, 0.20, 0.25\}$, $\nu = 0.6$, and $k$ was randomly picked from $4$ to $20$. The area covered by each dataset is very large, so to force objects to be reached by several sources, for each query, we picked source objects from the same cell (with the side equal to $d_{cc}$) at the beginning of the query interval. The results (see Fig.~\ref{TopKExper}) indicate that for top-k queries RICCmeetMin accesses in average about 37\% more pages than RICCtopK for the $MV$, and about 30\% more pages for $RW$  datasets. The advantage of RICCtopK owes to both, the RICCdecay index, 
and RICCtopK query processing. Information from the preprocessing allows for computing the maximum possible aggregate score $F_{max}$ using information from file {\em Reached(Hop)}, while RICCtopK reduces the number of objects that have to be accessed when the query reads the file {\em Meetings}.

\section{Conclusions}
\label{conclusion}
\vspace{-0.05in}
We presented two novel reachability problems: reachability with transfer decay and top-k reachability with transfer decay.
To process these queries efficiently, we designed two new algorithms: RICCdecay and RICCtopK. The RICCmeetMin algorithm~\cite{RICCmeet} was modified to answer the same types of queries, and served as a benchmark. We tested our algorithms on six realistic datasets, varying query duration and the maximum allowed number of hops. The performance comparison 
showed that RICCdecay and RICCtopK can answer the new types of queries more efficiently than RICCmeetMin.
 \vspace{-0.25em} 


  \vspace{-0.05in}
\bibliographystyle{IEEEtran}
\bibliography{bibfile.bib}


\end{document}